\documentclass[%
aps,
 amsmath,amssymb,
 reprint,%
]{revtex4-2}

\usepackage{graphicx,epsfig,epsf,float,amsmath,amssymb}
\usepackage{graphicx}
\usepackage{dcolumn}
\usepackage{bm}
\usepackage{hyperref}
\usepackage{subcaption}
\usepackage{braket}
\usepackage{scrextend}
\usepackage{float}
\usepackage{natbib}

\usepackage{amsfonts,stmaryrd} 

\usepackage{enumerate} 

\usepackage[ruled]{algorithm2e} 

\usepackage[framemethod=tikz]{mdframed} 

\usepackage{listings} 

\usepackage{multirow}
\usepackage{color, colortbl}
\usepackage{makecell}
\usepackage{mathtools}

\usepackage{fontenc}
\usepackage{graphicx}
\usepackage{subcaption}
\usepackage{XCharter} 

\def\v#1{{\bf#1}}
\def\mb#1{{\mathbf#1}}
\def\be{\begin{equation}}
\def\ee{\end{equation}}
\def\ben{\begin{equation*}}
\def\een{\end{equation*}}
\def\bea{\begin{eqnarray}}
\def\eea{\end{eqnarray}}

\newcommand{\cet}[1]{|#1)}
\newcommand{\cra}[1]{(#1|}

\def\<{\langle}
\def\>{\rangle}
\newcommand{\unit}{1\!\!1}

\begin{document}

\title{An exactly solvable tight-binding billiard in graphene}
\author{D. Condado  and E. Sadurní}
\affiliation{Instituto de Física, Benemérita Universisdad Autónoma de Puebla, Apartado Postal J-48, 72570. Puebla, México}
\date{December 2024}

\begin{abstract}
A triangular graphenic billiard is defined as a planar carbon polymer in the Hückeloid approximation of $\pi-$band electrons. It is shown that the equilateral triangle of arbitrary size and zig-zag edges allows for exact solutions of the associated spectral problem. This is done by a construction of wave superpositions similar to the Lamé solution of the Helmholtz equation in a triangular cavity, revisited by Pinsky. Exact wave functions, eigenvalues, degeneracies, and edge states are provided. The edge states are also obtained by a non-periodic construction of waves with vanishing energy. A comment on its connection with recent molecular models, such as triangulene, is given.    
\end{abstract}

\maketitle

\section{Introduction}

The interest in the study of graphenic structures \cite{Geim, Katsnelson} has led the research focus to different low-dimensional realizations such as nanotubes \cite{review} wound in various topologies \cite{micrograph}, as well as nanoribbons \cite{nanoribbons} together with their interesting electronic transport properties. Recently, smaller structures such as cycloacenes \cite{hiddendual,electronphonon} have been shown to possess exact solutions associated with their Hückeloid model, and from the experimental point of view, the fluorescent spectrum of cycloparafenylenes \cite{MCPP} has been shown to depend strongly on the size and topology of polymers. Some molecules recently known as \textit{coronene} \cite{Kleincor} and \textit{triangulene} \cite{Kleintri} have also attracted considerable attention due to their simplicity and promising applications of their corresponding edge states. The latter example, triangulene, closely resembles a triangular billiard with Dirichlet conditions made of a graphene flake, except for some extra carbon sites at the triangle vertices; these special polymers may actually appear in concrete realizations if cut from a honeycomb sheet and saturated with two hydrogen atoms at the three carbon vertices.

In addition to this important motivation in nanoscopic physics, it should also be mentioned that Dirac billiards -specifically, Weyl neutrino billiards \cite{neutrinobilliards}- have not encountered so far a concrete realization in particle physics. With the advent of graphene, there seems to be a chance to observe some interesting relativistic properties such as Klein tunneling \cite{Kleintun}, \textit{Zitterbewegung} \cite{Zitterbewegung,SetareMajari} and some aspects of MIT bag models of quark confinement \cite{bag1,bag2,bag3}. Indeed, the billiard can be made of a honeycomb carbon lattice, but we are about to see that the edge states emerging from this geometry cannot be built from Dirac spinors, and deserve a separate discussion.  

Paper structure: In this work, we solve the stationary Hückeloid model analytically for an equilateral triangle made of a honeycomb structure. Before proceeding to the solution, some mathematical remarks concerning the method of segmentation in polymers \cite{Hall, eigenpersistence} and tessellation in billiards are in order (Section 2). Then we proceed in two parts: First, we impose Dirichlet conditions exclusively along lines within one triangular sublattice and find all quantized energies (Section 3). Then, we tackle the problem of edge states, their orthogonalization and their $C_{3v}$ representations (Section 4). Conclusions are drawn in Section 6.  

\section{Bloch waves and eigen-persistence}

Eigen-persistence is a general phenomenon in graph theory that affects large Hückeloid structures \cite{Huckel1,Huckel2} when separated into smaller parts that possess eigenvalues in common with the parent (larger) structure. This is explained by nodal domains of the wavefunctions pertaining to the parent system. This phenomenon appears both in presence and absence of configurational symmetry; notably, it applies for Hamiltonians with no apparent symmetry of reflective or translational nature. In the case at hand,
the infinite graphene sheet is periodic, but the triangular cut is not, and it retains only a $C_{3v}$ subgroup. This makes our task more interesting in connection with equilateral triangular domains.

First, we denote by $H_c, H_e, H_n$ the tight-binding Hamiltonians of the central, external, and nodal domains, respectively. As they correspond to different lattice sites, they constitute a direct sum decomposition of the full Hamiltonian $H$ of the infinite sheet. These three summands are coupled by adjacency matrices $\Delta_{cn}, \Delta_{en}$, but not $\Delta_{ce}$, as these would involve second-neighbor interactions beyond Hückel's approximation. Thus, we have
\begin{gather}
\label{1}
H=\left(\begin{array}{ccc}
H_e & \Delta_{en} & 0 \\
\Delta_{en}^\dagger & H_n & \Delta_{cn} \\
0 & \Delta_{cn}^\dagger & H_c
\end{array}\right),\\
\label{2}
H\ket\psi=E\ket\psi,\quad H_i\ket{\psi_i}=E\ket{\psi_i},\quad i=e,c,\\
\label{3}
\ket\psi=\left(\begin{array}{c}\ket{\psi_e} \\ 0 \\ \ket{\psi_c}\end{array}\right),\quad \Delta_{en}^\dagger\ket{\psi_e}+\Delta_{cn}\ket{\psi_c}=0.
\end{gather}
Here we note that in continuous variables, and in the presence of a nodal line that imposes antisymmetry, (\ref{3}) reduces to
\be
\ket\psi=\left(\begin{array}{c}\ket\phi \\ 0 \\ -\ket\phi\end{array}\right),\quad\ket{\psi_e}=\ket\phi=-\ket{\psi_c}.
\ee
We see an effective disconnection of regions $e, c$, but not necessarily an antisymmetry of $|\psi \rangle$ with respect to the nodal line where $|\psi \rangle = 0$. For lattices that do possess reflection symmetry around such nodal domains, the task is easier; in the continuous limit, the Helmholtz equation subjected to Dirichlet conditions can be treated with this trick, which leads to full space tessellations \cite{fruits, sunada} as originally seen by Lamé \cite{historical1, historical2}. Here, on the other hand, we have $H_e \neq H_c$ for adjacent cells. This is illustrated in fig. \ref{latticevectors} and \ref{barbado}, and it defines a fundamental cell made of two triangular polymers (a parallelogram) that fills the infinite honeycomb structure. Important implications from this observation will be seen throughout this paper: The emergence of periodic solutions whose nodal domains produce the desired polymer and non-periodic solutions that correspond to edge states.

In passing, we note that not all discrete problems have the two alluded types of solution. A typical example in the opposite direction is a finite chain, viewed as a cell with Dirichlet conditions inside an infinite row of atoms. Sinusoidal waves $\phi_n = \sin\pi n/N$ solve the tight-binding recursion $\phi_{n+1} + \phi_{n-1} = E \phi_n$ with a finite spectrum $E_n = 2 \cos(\pi n/N), n=1,...,N$. They also solve the infinite problem, as translations by a cell imply $\sin(\pi n/N) = - \sin(\pi(n+N)/N)$, i.e. $H_e = H_c$ and the antisymmetry $|\psi_e \rangle = - |\psi_c \rangle $ is fulfilled, together with $\phi_N = 0$. For graphene, on the other hand, a triangular tessellation of alternating signs is not possible, but a real Bloch wave that covers both $H_e$ and $H_c$ coupled across linear boundaries (see fig. \ref{barbado}) is constructible, and the additional relation $\Delta \psi = - \Delta \psi$ is not necessary.  

\section{Triangular billiards from vanishing upper components of Bloch spinors}

\begin{figure}
  \centering
      \includegraphics[width=0.8\linewidth]{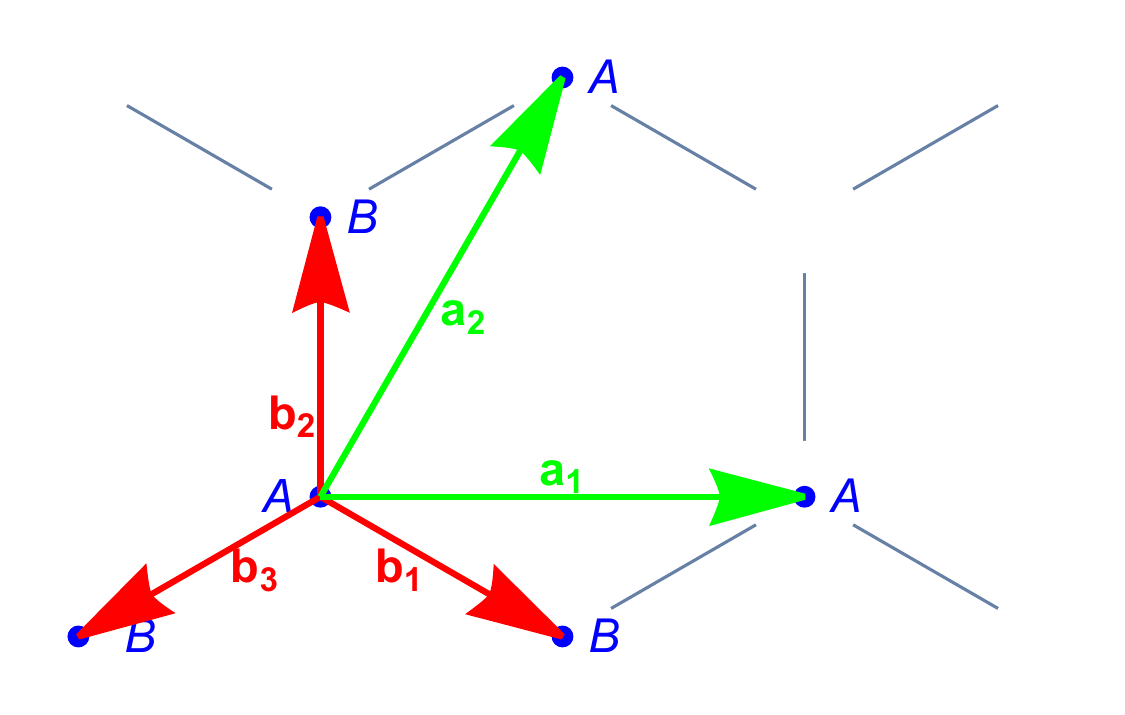}
     \caption{Primitive vectors $\{ \mb b_1, \mb b_2, \mb b_3\}$ of hexagonal lattices in this paper. The set $\{\mb a_1, \mb a_2\}$ spans subtriangular lattices $A$ or $B$, with integers $n_1$ and $n_2$, as shown in (\ref{sites_pos}).}
     \label{latticevectors}
\end{figure}

A full tight-binding model for electrons hopping on an infinite graphene sheet with translated Fermi energy $E_0=0$ and unit hopping amplitude $\Delta=1$ is described by a nearest-neighbor Hamiltonian $H$ given by \cite{playing}
\begin{equation}
    \label{ketbraH}
H=-\sum_{n_1,n_2}\sum_{i=1,2,3}\ket{\mb A+\mb b_i}\bra{\mb A}+\text{H.c.},
\end{equation}
with the usual parameterizations (see fig. \ref{latticevectors})
\begin{gather}
\label{sites_pos}
\mb A=n_1\mb a_1+n_2\mb a_2,\quad \mb a_1=\hat\imath,\quad\mb a_2=\frac12(\hat\imath+\sqrt3\hat\jmath),\\
\mb b_1=\frac1{2\sqrt3}(\sqrt3\hat\imath-\hat\jmath),\quad\mb b_2=\frac1{\sqrt3}\hat\jmath,\quad \mb b_3=-\mb b_1-\mb b_2.
\label{2}    
\end{gather}

The waves propagating on this sheet can be written in the form $ | \psi \rangle = C_A | \psi_A \rangle + C_B | \psi_B \rangle$, i.e. a bipartite decomposition using $A$-$B$ spinors.

\begin{figure}
  \centering
      \includegraphics[width=0.8\linewidth]{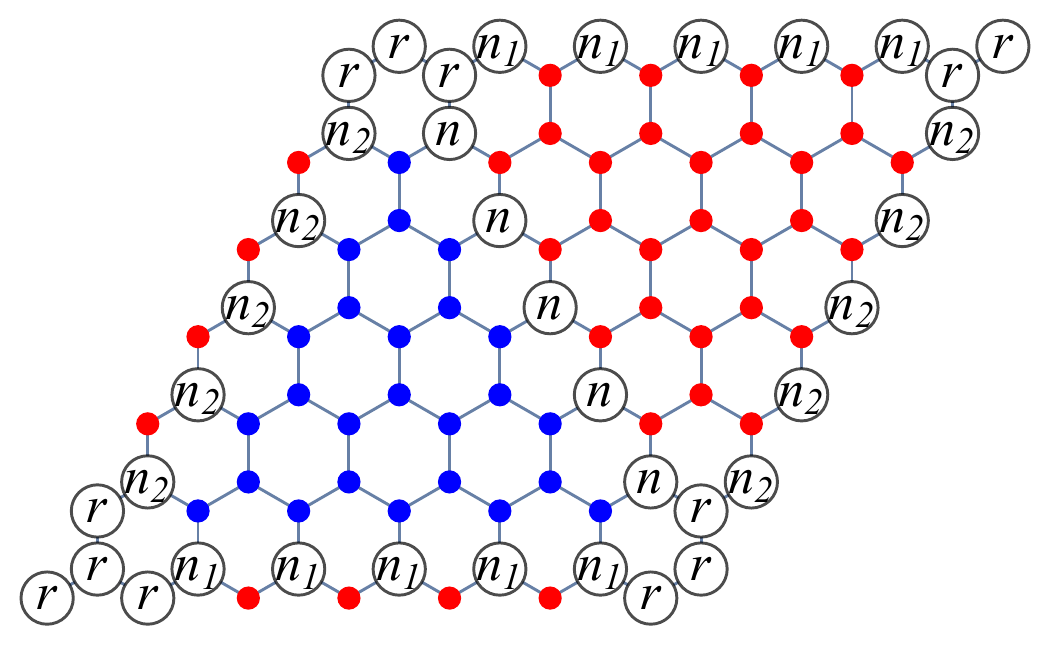}
     \caption{Nodal requirements isolating triangulene segments (zig-zag: blue, armchair: red) in an infinite sheet. The nodal lines with labels $n_1$, $n_2$ and $n$ are imposed in (\ref{requirements}). Nodes labeled by $r$ are implied by the H\"uckel recurrence. In this example, $N=6$.}
     \label{barbado}
\end{figure}

A triangular billiard resembling a carbon polymer with active sites at the vertices (also called \textit{triangulene}, but adding a carbon site at each vertex) can be described in this notation using (\ref{ketbraH}) with additional restrictions for site numbers $n_1,n_2$, which are finite. Alternatively, the system can be defined by a sheet with nodal lines along horizontal and $\pm 60$-degree directions, as shown in fig. \ref{barbado}. 

The solutions of the tight-binding problem associated with the stationary Schr\"odinger equation can be written analytically by combining plane waves. With the aid of the complex quantity
\begin{align}
\alpha=e^{-i\mb k\cdot\mb b_2}\sum_{i=1}^3e^{i\mb k\cdot\mb b_i}=1+e^{-i\mb k\cdot(\mb a_2-\mb a_1)}+e^{-i\mb k\cdot\mb a_2},
\end{align}
together with canonical vectors 
\begin{equation}
    |A\rangle= (1,0)^{\rm T},\quad|B\rangle = (0,1)^{\rm T},
\end{equation}
one has
\begin{gather}
\label{solutions}
\psi_k^{(\pm)}=\frac1{\sqrt2}\left(\begin{array}{c}1\\\pm\phi_k\end{array}\right)e^{i\mb k\cdot\mb A},\quad H_k=\left(\begin{array}{cc}0& \alpha^*\\\alpha&0\end{array}\right),\\
H_k\psi_k^{(\pm)}=\pm E_k\psi_k^{(\pm)},\quad E_k=|\alpha|,\quad\varphi_k=\arg\alpha.
\end{gather}

It should be noted that our fundamental dimer is chosen in the direction of $\mb b_2$, i.e. vertically displayed in fig. \ref{latticevectors}. The translation along this vector is given by the swap operator $\sigma_x$ in the canonical basis. This standard notation makes apparent the difference between linear combinations of $A$-waves and $B$-waves due to the presence of the $k$-dependent coefficient $\phi_k$.

\begin{figure}
  \centering
      \includegraphics[width=0.6\linewidth]{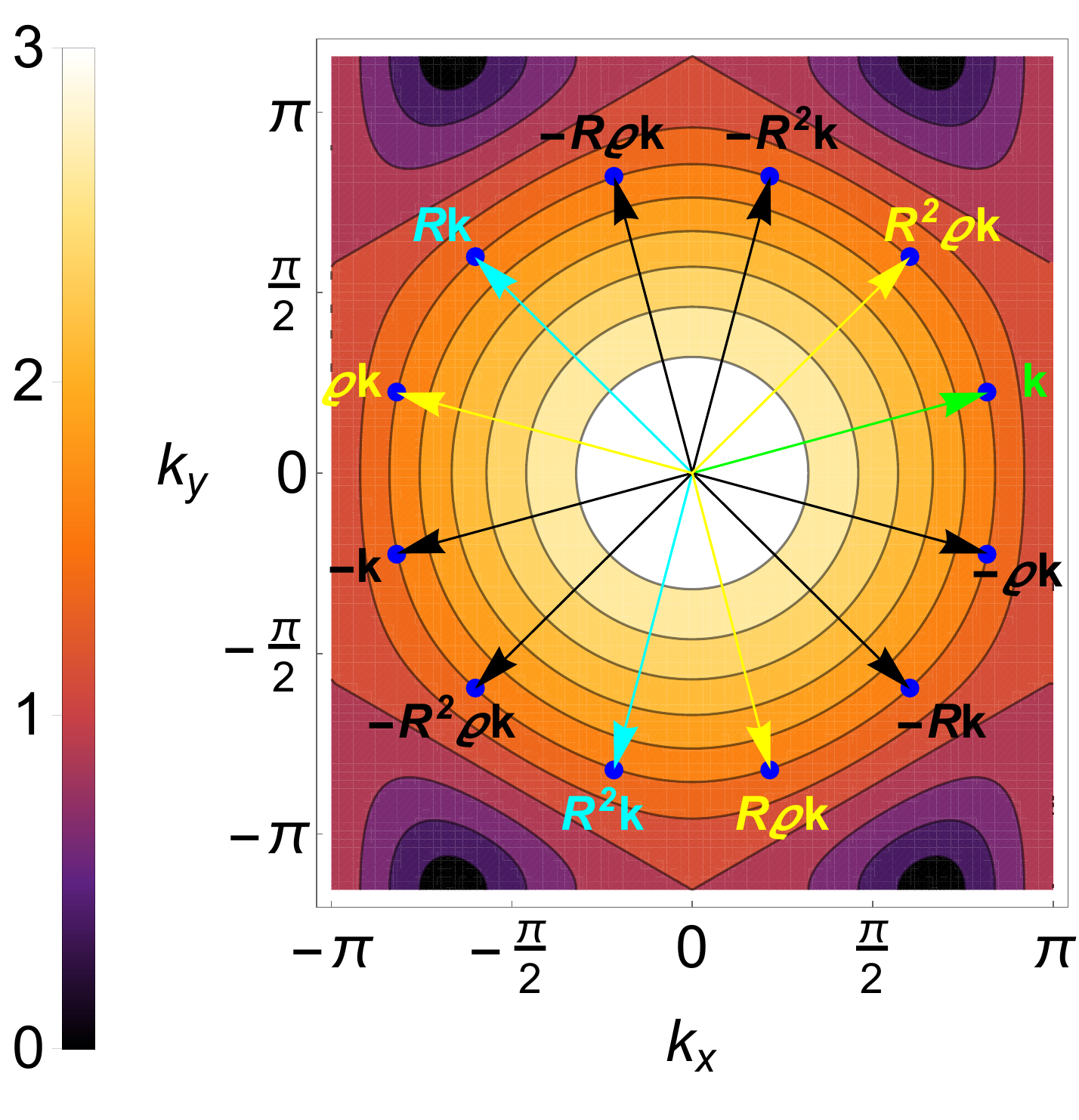}
     \caption{The contours (level curves) for the graphene spectrum (\ref{spectrum}): $E_+(k_x,k_y)$. The arrows represent the 12 degenerate solutions presented in (\ref{12}) and the blue points indicate the degenerate energies.}
     \label{contours}
\end{figure}

In periodic lattices such as the honeycomb configuration, not all vectors $\v k$ in the first Brillouin zone $\Omega_{\rm B}$ represent different states (an interesting consequence of this is the so-called \textit{umklapp} effect \cite{umklapp}). The values of $\v k$ connected by a vector $\v k_l = 4\pi(m_1 \v a_1 + m_2 \v a_2)$ belonging to a triangular sublattice of the full reciprocal space represent the same state, i.e. the spinor is invariant under such translations:

\begin{align}
\psi_{k+k_l}=\psi_k,\quad\alpha_{k+k_l}=\alpha_k,\quad E_{k+k_l}=E_k.
\end{align}
In addition, a vector $\v k_r = 4\pi q\v b_2 = 4\pi q \v j /\sqrt{3}, q=0, \pm1, \pm2,... $, connecting the two reciprocal sublattices and producing a hexagonal lattice in $\v k$ space, leads to the same energy as above, but modifies the spinor. Similarly, the inverted vector $-\v k$ is equivalent to complex conjugation and thus leads to a linearly independent spinor:

\begin{align}
\psi_{-k}=\psi_k^*=\frac1{\sqrt2}\left(\begin{array}{c}1\\\phi_k^*\end{array}\right)e^{-i\mb k\cdot\mb A}.
\end{align}
From these considerations, we can see that the relation $|\alpha_k | = |E_k|$ is invariant under the group $C_{6v}$ applied on the hexagon of $\Omega_{\rm B}$ around $\v k=0$, and assuming that $\v k \neq \v k_D$ (the famous Dirac point \cite{eladDirac}), there are, in general, 12 vectors $\v k_T$ that lead to the same $|E_k|$. The full list of vectors is generated by the application of $2 \pi/3$ rotations $R$, $4\pi/3$ rotations $R^2$, $x-$axis reflections $\varrho$ and inversions $\v k \mapsto -\v k$, as well as their compositions. The set is

\begin{align}
\label{12}
\v k_T \in \{ \pm \v k, \pm R \v k, \pm R^2 \v k, \pm \varrho \v k, \pm R \varrho \v k, \pm R^2 \varrho \v k \},   
\end{align}
see fig. \ref{contours}. 

Then, $\alpha_k$ and $\phi_k=\alpha_k /\left|\alpha_k\right|$ pick up new phases when group transformations are applied, according to
\be
\begin{gathered}
\varphi_{-k}=-\arg\alpha_k,\quad\varphi_{\varrho k}=\arg\alpha_k, \\
\varphi_{Rk}=-\mb k\cdot\mb a_2-\arg\alpha_k=\varphi_{R\varrho k}, \\
\varphi_{R^2 k}=\mb k\cdot\left(\mb a_2-\mb a_1\right)+\arg\alpha_k=\varphi_{R^2\varrho k},
\end{gathered}
\ee
similarly for the remaining six $\varphi_{\pm k_T}$. This allows to build combinations of real wave functions for the upper spinors with the three requirements that force nodal lines at the edges of the billiard (see fig. \ref{barbado}):
\be
\label{requirements}
\begin{gathered}
\sum_T a_T e^{i \mb k_T \cdot\mb a_1 n_1}=0, \text { for $n_2=0$ and $n_1=1,\dots,N$,} \\
\sum_T a_T e^{i \mb k_T \cdot\mb a_2 n_2}=0, \text { for $n_1=0$ and $n_2=1,\dots,N$,} \\
\sum_T a_T e^{i \mb k_T \cdot\left(n\mb a_1+[N-n]\mb a_2\right)}=0, \text { for $n=1,\dots,N$.}
\end{gathered}
\ee

Here, $a_T$ are the coefficients of the superposition and $N$ is the semi period or distance between nodal lines, i.e. the number of sites on the side of the triangle plus one (counted along sites B, indicated in fig. \ref{latticevectors}). These equations are solved as follows.

The linearly independent functions of $n_1$ or $n_2$ in the first two equations may vanish only if all the exponential functions have the same amplitude, and thus $\left|a_T\right|=\mathcal N$ is a common normalization factor. Then, to have real waves as solutions, one demands sinusoidal functions, built by each pair $\pm\mb k_T$ and so $a_T=i\mathcal N e^{i \alpha_T},$ $a_{-T}=a_T^*$. From 12 terms in (\ref{requirements}) we are down to 6:
\be
\label{condition for 12}
\sum_{r=1}^6 \sin \left(\mb k_r\cdot\mb a_in_i+\alpha_r\right)=0, \quad i=1,2,
\ee
where $\mb k_r\in\left\{\mb k,\varrho\mb k,R\mb k ,R\varrho\mb k ,R^2\mb k,R^2\varrho\mb k \right\}$. If $i=1$, substitution of $\mb k_r\cdot\mb a_1 n_1$ shows that the terms are pairwise canceled if $\alpha_1=-\alpha_2,$ $\alpha_3=-\alpha_4,$ $\alpha_5=-\alpha_6$. If $i=2$ one has $\alpha_1=\alpha_3=\alpha_5$ for another pairwise cancellation. The energy quantization condition emerges from the third relation in (\ref{requirements}) involving $N$, as it closes the cavity and restricts the solutions to a finite number of vectors (one has $N(N-1)/2+(N-1)(N-2)/2=(N-1)^2$ points inside the polymer).

\begin{figure*}
  \centering
      \includegraphics[width=0.8\linewidth]{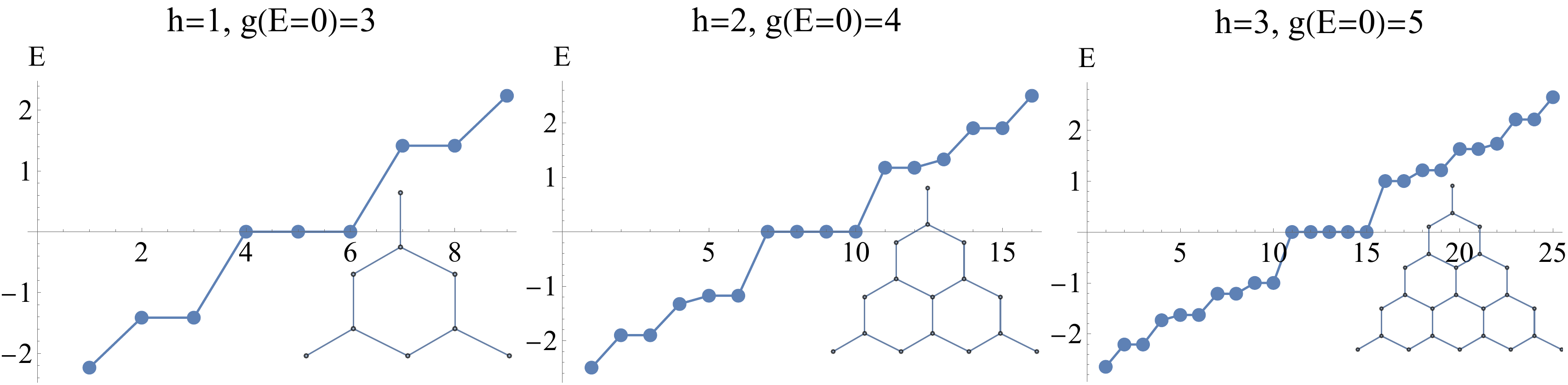}
     \caption{Numerical spectra of triangulene molecules with increasing size, $h=N-3$ is the number of hexagonal rings at the baseline, $g(E)$ denotes the degeneracy. It is clear that $g(0)$ increases linearly with the molecular size.}
     \label{numeric}
\end{figure*}

The third condition leads us to a trigonometric Diophantine equation (see Appendix \ref{diof}) that is solved for all values of $n$ by the following quantization conditions,
\begin{gather}
\label{k quantx}
\mb k \cdot\mb a_1=2 \pi(2q+p)/ 3 N,\\
\label{k quanty}
\mb k \cdot\mb a_2=2 \pi(2p+q)/ 3 N,\\
\mb k=4 \pi\left(q\,\mb a_1+p\,\mb a_2\right)/3N,
\end{gather}
where $q, p$ are integers, so far unconstrained. In Cartesian components, we have
\be
k_x=2 \pi(2q+p)/ 3 N, \quad k_y=2\pi p/\sqrt3N.
\ee
Direct substitution of $(k_x,k_y)$ into graphene's dispersion relation gives a formula for quantized energies
\be
\begin{aligned}
\label{spectrum}
&E_{\pm}(q,p)=\\
&\pm \sqrt{3+2 \cos k_x + 2 \cos \frac{k_x + \sqrt{3}k_y }{2}+2 \cos \frac{k_x - \sqrt{3}k_y }{2}}, 
\end{aligned}
\ee
and with the help of simple trigonometric identities we rewrite this expression as
\be
\begin{aligned}
\label{spectrum}
&E_{\pm}(q,p)=\pm \sqrt{1+4\cos^2\frac{k_x}2+4\cos\frac{k_x}2\cos\frac{\sqrt3k_y}2}\\
=&\pm \sqrt{1+4\cos^2\frac{\pi(2q+p)}{3N}+4\cos\frac{\pi(2q+p)}{3N}\cos\frac{\pi p}N}\\
=&\pm \sqrt{\sin^2\frac{\pi p}N+\left(2\cos\frac{\pi(2q+p)}{3N}+\cos\frac{\pi p}N\right)^2}.
\end{aligned}
\ee
The criterion for determining degeneracy comes from the set of indices $q,p$ that lead to a constant $E$. This is true even when $E=0$, but since this case is comprised of both periodic and aperiodic solutions, it must be analyzed separately. It should be noted that the band center accumulates an ever-growing number of states as a function of the size $N-1$; we show this in fig. \ref{numeric}, where the numerical spectra of three molecules are compared also as a function of their size; the number of hexagons or benzene rings $h$ at the baseline of the structure depends on $N$ as $h=N-3$.

\begin{figure}
  \centering       \includegraphics[width=0.8\linewidth]{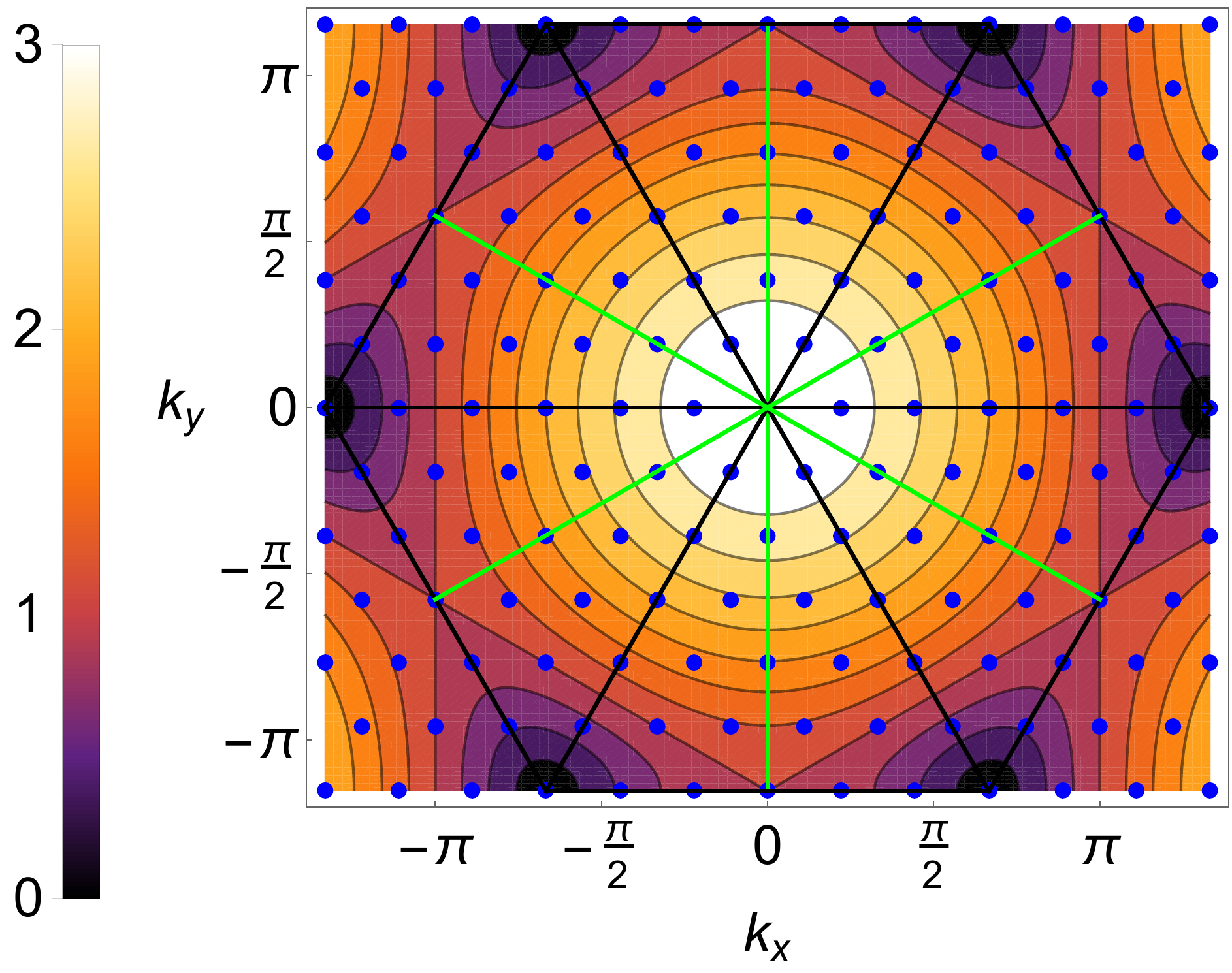}
     \caption{Spectrum for the $N=6$ triangulene. The blue dots represent the Pinsky lattice, i.e. all solutions according to (\ref{k quantx}) and (\ref{k quanty}). Only interior points in each triangle, including the green line, lead to l.i. solutions.}
     \label{smallt}
\end{figure}

Now we must pay attention to the restrictions obeyed by $q$ and $p$. In fig. \ref{contours} we show the 12 lattice points in reciprocal space employed in the construction of a single wave function. Along the symmetry lines of the hexagonal Brillouin zone described in fig. \ref{smallt} (black lines) the dispersion relation (\ref{spectrum}) is still valid, but the wave functions (\ref{requirements}) vanish identically. In order to properly account for a linearly independent (l.i.) set of eigenfunctions, we rule out $q,p$ such that $\v k = \hat\imath k$ or its images $\rho \v k, R \v k, R \rho \v k, R^2 \v k, R^2 \rho \v k$, i.e. those reciprocal vectors that lie within symmetry lines of the hexagon. Additionally, $q,p$ must be restricted to a single triangular sector within green and black lines in fig. \ref{smallt}; by convention, we choose the first triangular sector $0<k_y<\pi/\sqrt3$, $\sqrt3k_y\leq k_x<-k_y/\sqrt3+4\pi/3$. We note that the rightmost point at the boundary $k_y=0$, $k_x=4\pi/3$ is a \textit{Dirac point}. This translates into 
\be
\label{sector qp}
0<p<N/2,\quad p\leq q<N-p
\ee
and proper counting of l.i. states is now reduced to l.i. spinors, together with their degeneracies. The detailed spinorial analysis is presented in section \ref{completeness}. The full spinor for periodic solutions can now be written as 
\begin{widetext}
\be
\label{states c}
\begin{gathered}
\psi^{(\pm)}_{1,k}=\sqrt2\text{Re}[\Psi^{(\pm)}_k],\quad\psi^{(\pm)}_{2,k}=\sqrt2\text{Im}[\Psi^{(\pm)}_k],\\
\Psi^{(\pm)}_k=\frac1{\sqrt{12}}\Bigg\{
\left(\begin{array}{c}1\\\pm\phi_k\end{array}\right)e^{i\mb k\cdot\mb A}
+\left(\begin{array}{c}1\\\pm e^{i\mb k\cdot\mb a_2}\phi_k\end{array}\right)e^{-i\mb k\cdot\mb a_1(n_1+n_2)+i\mb k\cdot\mb a_2n_1}
+\left(\begin{array}{c}1\\\pm e^{i\mb k\cdot(\mb a_2-\mb a_1)}\phi_k\end{array}\right)e^{i\mb k\cdot\mb a_1n_2-i\mb k\cdot\mb a_2(n_1+n_2)}\\
-\left(\begin{array}{c}1\\\pm\phi_k^*\end{array}\right)e^{-i\mb k\cdot\mb a_1n_2-i\mb k\cdot\mb a_2n_1}
-\left(\begin{array}{c}1\\\pm e^{-i\mb k\cdot\mb a_2}\phi_k^*\end{array}\right)e^{i\mb k\cdot\mb a_1(n_1+n_2)-i\mb k\cdot\mb a_2n_2}
-\left(\begin{array}{c}1\\\pm e^{i\mb k\cdot(\mb a_1-\mb a_2)}\phi_k^*\end{array}\right)e^{-i\mb k\cdot\mb a_1n_1+i\mb k\cdot\mb a_2(n_1+n_2)}
\Bigg\}.\\
\end{gathered}
\ee
\end{widetext}

\begin{figure}
  \centering
      \includegraphics[width=\linewidth]{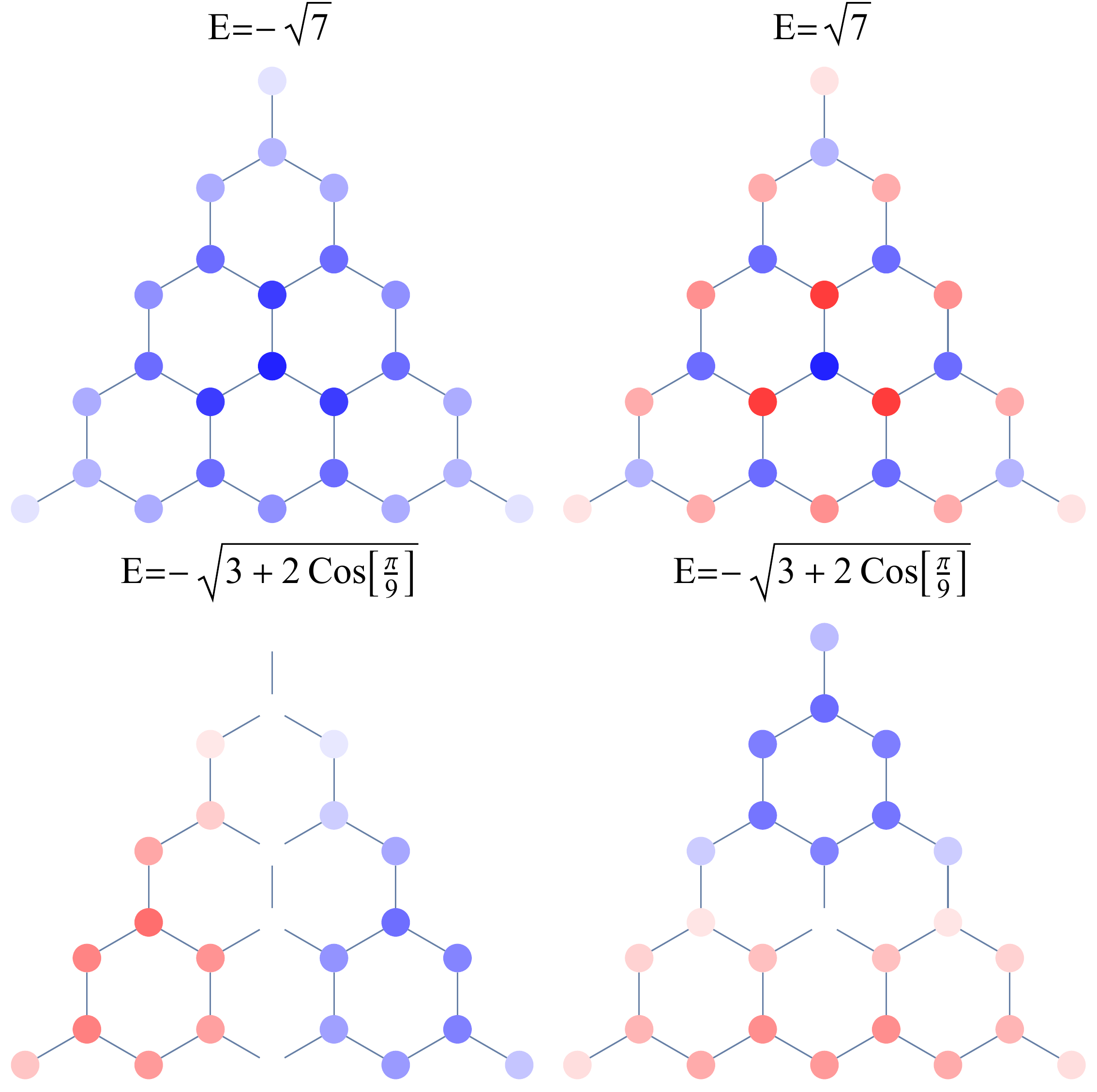}
     \caption{Wave functions of triangulene $N=6$ according to (\ref{states c}). Top left panel: ground state. Top right:  most excited state. Bottom: two degenerate states, $\psi_1$ and $\psi_2$ in eq. (\ref{states c}).}
     \label{repre}
\end{figure}

These solutions contemplate all interior points $q,p$ in a fundamental cell of \textit{periodic solutions} of triangulene. In the next section, we investigate \textit{edge states} separately. The construction described above closely resembles Pinsky's lattice \cite{Pinsky} for the triangular billiard, but the existence of edge states is exclusive to discrete billiards. In our case, the allowed points belong to contours of constant energy $E=\left|\alpha_k\right|$, which are not circles, in contrast with the Helmholtz case. Once again we resort to fig. \ref{contours} to illustrate these properties. We close this section by showing in fig. \ref{repre} four representative waves in (\ref{states c}) for $N=6$.

\section{Solution subspace of edge states}

The presence of edge states \cite{Bellec, diracpointedgestates} in the solutions of triangular billiards represents an important feature in the electronic transport of finite graphene \cite{Nakada,gopar}. As we saw in the previous section, such states cannot be reproduced using the chiral spinors corresponding to the Dirac point in the full honeycomb lattice, despite the fact that $E=0$ for these solutions. Moreover, such states grow linearly in number with the size of the polymer; therefore, their presence cannot be neglected. Since the wave function must be constructed in a non-periodic manner, superposition of graphene eigenstates is no longer useful, and we must find the edge states $\ket e$ from scratch. We begin by writing them in a general fashion:
\be
\label{edge definition}
\begin{gathered}
\ket e=\sum_{n_1,n_2}(a_{n_1,n_2}\ket{\mb A}+b_{n_1,n_2}\ket{\mb A+\mb b_2}),\quad H\ket e=0,\\
n_2=0,\dots,N-2,\quad n_1=1,\dots,N-1-n_2,\quad a_{n_1,0}=0,
\end{gathered}
\ee
where indices $n_1,n_2$ are the same as in (\ref{sites_pos}) and $a_{n_1,0}=0$ is due to the nodal line. After substitution into (\ref{ketbraH}), we arrive at the recurrence relations:
\begin{gather}
\label{pre a}
a_{n_1-1,n_2+1}+a_{n_1,n_2+1}+a_{n_1,n_2}=0 \times b_{n_1,n_2},\\
\label{pre b}
b_{n_1+1,n_2-1}+b_{n_1,n_2-1}+b_{n_1,n_2}=0 \times a_{n_1,n_2},
\end{gather}
where the zeros on the right-hand side are due to $E=0$. Note that in (\ref{pre a}) we need to employ the remaining nodal lines given by 
\be
\begin{gathered}
a_{0,n_2}=0,\quad n_2=0,\dots,N,\\
a_{n,N-n}=0,\quad n=0,\dots,N.
\end{gathered}
\ee
With these considerations, we can prove that $a_{n_1,n_2}=0$ altogether. We begin by noting that $a_{1,1}=0$ and by successive use of (\ref{pre a}), we have $a_{n_1,1}=0$ and $a_{1,n_2}=0$. We can repeat the process starting with $a_{2,2}=0$ and then again with $a_{3,3}=0$ until we have covered all the amplitudes.
This leaves us only with one recurrence, namely
\be
\label{rec b}
\begin{gathered}
b_{n_1+1,n_2-1}+b_{n_1,n_2-1}+b_{n_1,n_2}=0,\\
n_2=1,\dots,N-2,\quad n_1=1,\dots,N-1-n_2.
\end{gathered}
\ee
No further subcases are needed here.


This expression, (\ref{rec b}), is the building block for all l.i. edge states. As a last remark before their construction, if we consider all the relations in recurrence (\ref{rec b}), we have a total of $(1/2)N(N-1)$ free parameters versus $(1/2)(N-1)(N-2)$ equations, which shows that the total number of l.i. edge states is
\be
\frac12N(N-1)-\frac12(N-1)(N-2)=N-1.
\ee
This will be employed in the following.


\subsection{Edge states basis as a representation of $C_{3v}$}

We begin by defining $R$ as the operator that rotates a given eigenstate counterclockwise by $2\pi/3$
. We can employ $R$ to construct a polynomial projection operator
\be
\label{Urotation}
\Pi_q=\frac19(1+z^{-q}R+z^qR^2),\quad z=e^{i2\pi/3},\quad q=-1,0,1.
\ee
The following properties are easy to deduce
\be
\label{C3vsym}
R\Pi_q=z^q\Pi_q,\quad\Pi_p^\dagger \Pi_q=\delta_{q,p}\Pi_q,
\ee
from which we conclude that the subspace $\Pi_q$ projects the edge states into the eigenbasis of the $C_{3v}$ group, which commutes with the Hamiltonian. These attributes 
allow us to infer that from an arbitrary edge state without $C_{3v}$ symmetry $\ket e$ we can generate a set of 3 orthogonal edge states $\Pi_q\ket e$ with $C_{3v}$ symmetry. The starting states $\ket e$'s —from now referred to as seed states— should be chosen as linearly independent. For this purpose, we introduce a construction based on layers.
\begin{figure}
  \centering
      \includegraphics[width=0.4\linewidth]{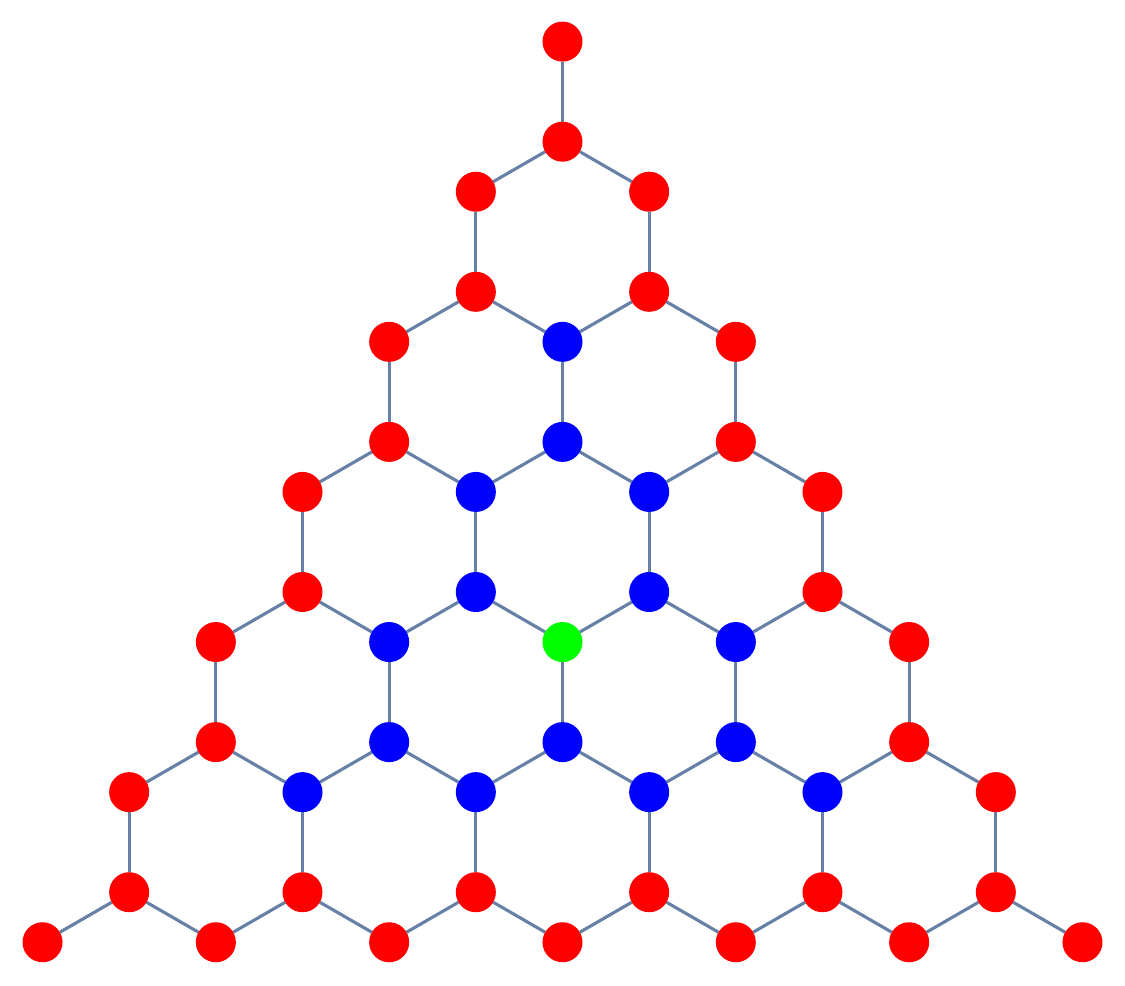}
     \caption{The $N=8$ triangulene has three layers, red is layer 1, blue is layer 2 and green is layer 3.}
     \label{layers0}
\end{figure}
\begin{figure}
  \centering
    \includegraphics[height=3.5cm]{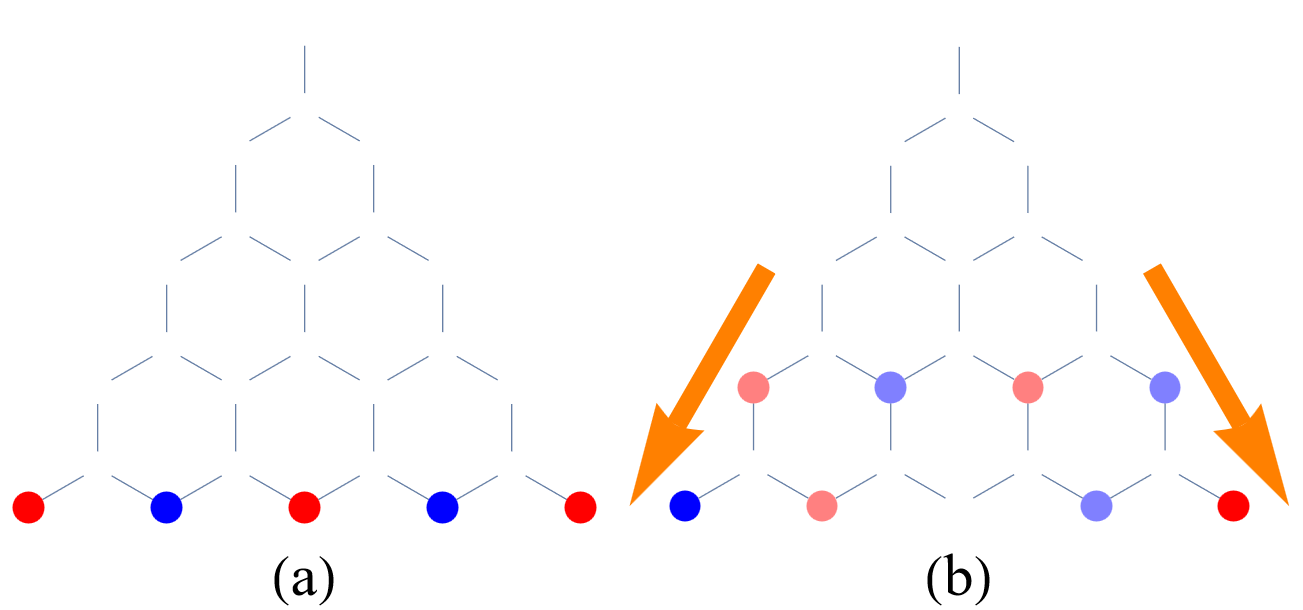}
   \caption{Seed states $\ket l$ of the $N=6$ triangulene. The orange arrows indicate the generation of amplitudes consistently with recurrence (\ref{rec b}); amplitudes of alternating signs in rows $l$ (counting from the bottom) cancel the rows above them. (a) Bottom seed state $\ket1$. (b) If $l>1$, the rows below row $l$ are filled in a antisymmetric (symmetric) manner if $N+l$ is even (odd), leaving nodes at central sites, such that the number of sites with non-vanishing amplitudes is the same in all rows.}
  \label{bothlayers}
\end{figure}

We define a layer of a triangular polymer as all the sites within a triangular loop (see fig. \ref{layers0}). We count them from the outermost to the innermost. 
It is clear that the number of layers $N_L$ in a triangular polymer of size $N$ is \linebreak $\lceil (N-1)/3\rceil$, i.e., the upper integer part of $(N-1)/3$. 

We need a total of $N_L$ seed states, which we build according to alternating signs in the amplitudes of a horizontal row and proceed downward by employing the recurrence (\ref{rec b}); by construction, this constitutes a solution of such a recurrence. A graphical explanation is provided in figs. \ref{bothlayers}a-\ref{bothlayers}b. In this fashion, we can generate $3(N_L-1)$ edge states compatible with the representations of $C_{3v}$ via

\be
\label{non ort edge states}
\begin{gathered}
\ket{l,q}\equiv\Pi_q\ket l,\\
q=-1,0,1,\quad l=1,\dots,N_L-1,\quad N_L \equiv \lceil (N-1)/3\rceil.
\end{gathered}
\ee
We should note that, by construction, these $3(N_L-1)$ states are l.i. for different values of $l$ and due to (\ref{C3vsym}), they are orthogonal for different $q$. We must now verify that there are no more l.i. edge states that can be constructed without filling the innermost layer. This layer consists of a single central point if $3N_L=N+1$, for which the wave does not vanish, and three points arranged in a star if $3N_L=N$ for which the central point corresponds to an $A$-site. The completeness of the l.i. set is done by considering that edge states with vanishing amplitudes in the innermost layer have $(3/2))(N_L-1)(2N-3N_L+2)$ unknown amplitudes, while (\ref{rec b}) provides $(3/2))(N_L-1)(2N-3N_L)$ equations, which leaves us with no more than $3(N_L-1)$ free parameters. This reasoning proves that so far we have only obtained $3(N_L-1)$ l.i. states, the remaining $N-1-3(N_L-1)$ edge states are constructed according to the following three cases:
\begin{figure}
  \centering
      \includegraphics[width=0.3\linewidth]{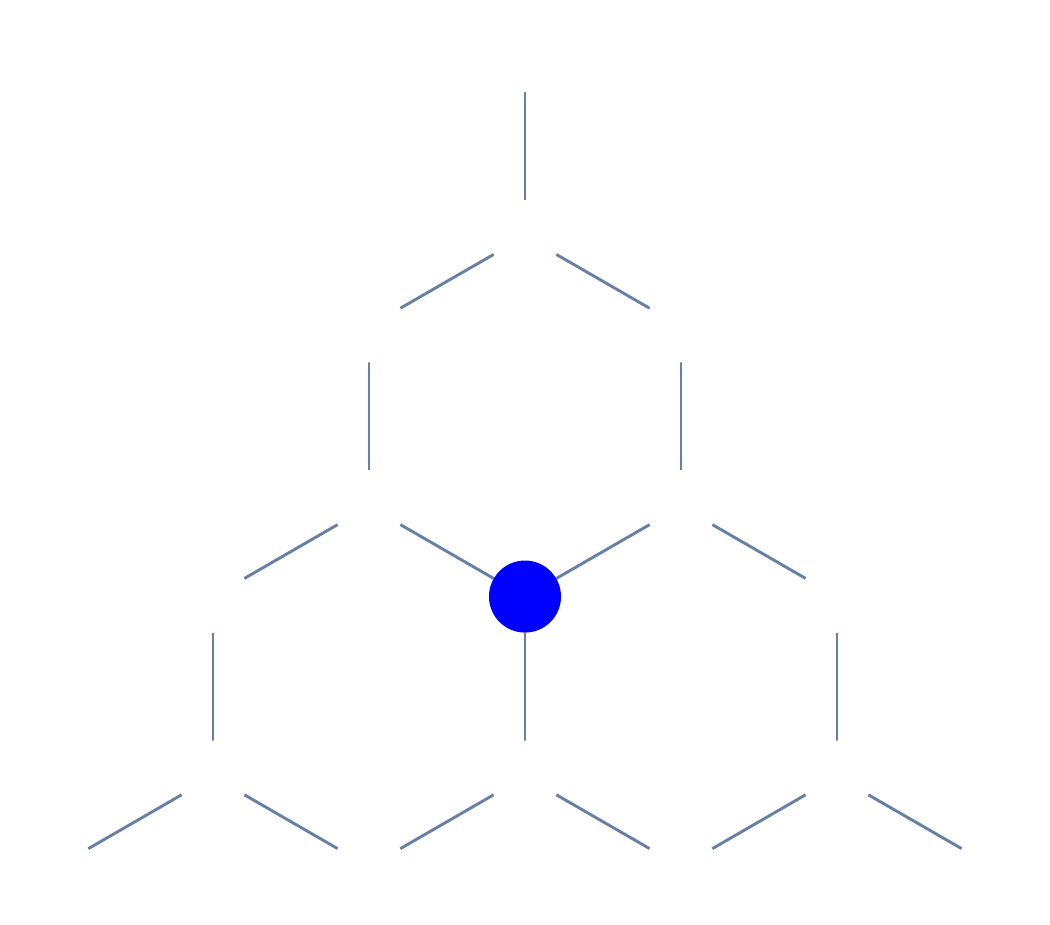}
     \caption{A fixed point in a triangulene molecule ($N=5$). This point is left invariant under rotations and only exists in triangulenes such that $3N_L=N+1$.}
     \label{central}
\end{figure}

\begin{figure}[ht]
  \centering
    \includegraphics[width=\linewidth]{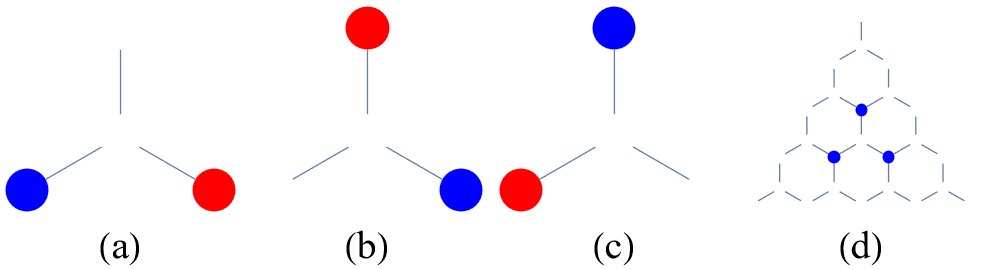}
  \caption{Sites in the innermost layer with $3N_L=N$. (a) For $N=3$, the seed state $\ket1$ is the antisymmetric edge state. (b) Counter clockwise rotation $R\ket1$. (c) Clockwise rotation $R^\dagger\ket1$. (d) The $N=3$ structure embedded in a larger polymer ($N=6$).}
  \label{N2layer}
\end{figure}

\begin{figure*}[ht]
  \centering
    \includegraphics[width=\linewidth]{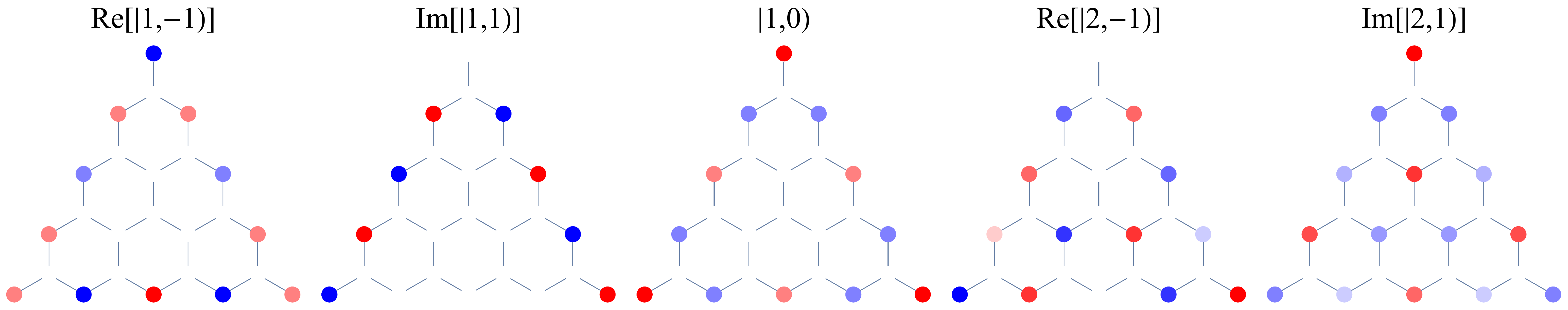}
  \caption{The 5 real orthonormal edge states of the $N=6$ triangulene.}
  \label{ort ed state}
\end{figure*}

\begin{figure}[ht]
  \centering       \includegraphics[width=0.8\linewidth]{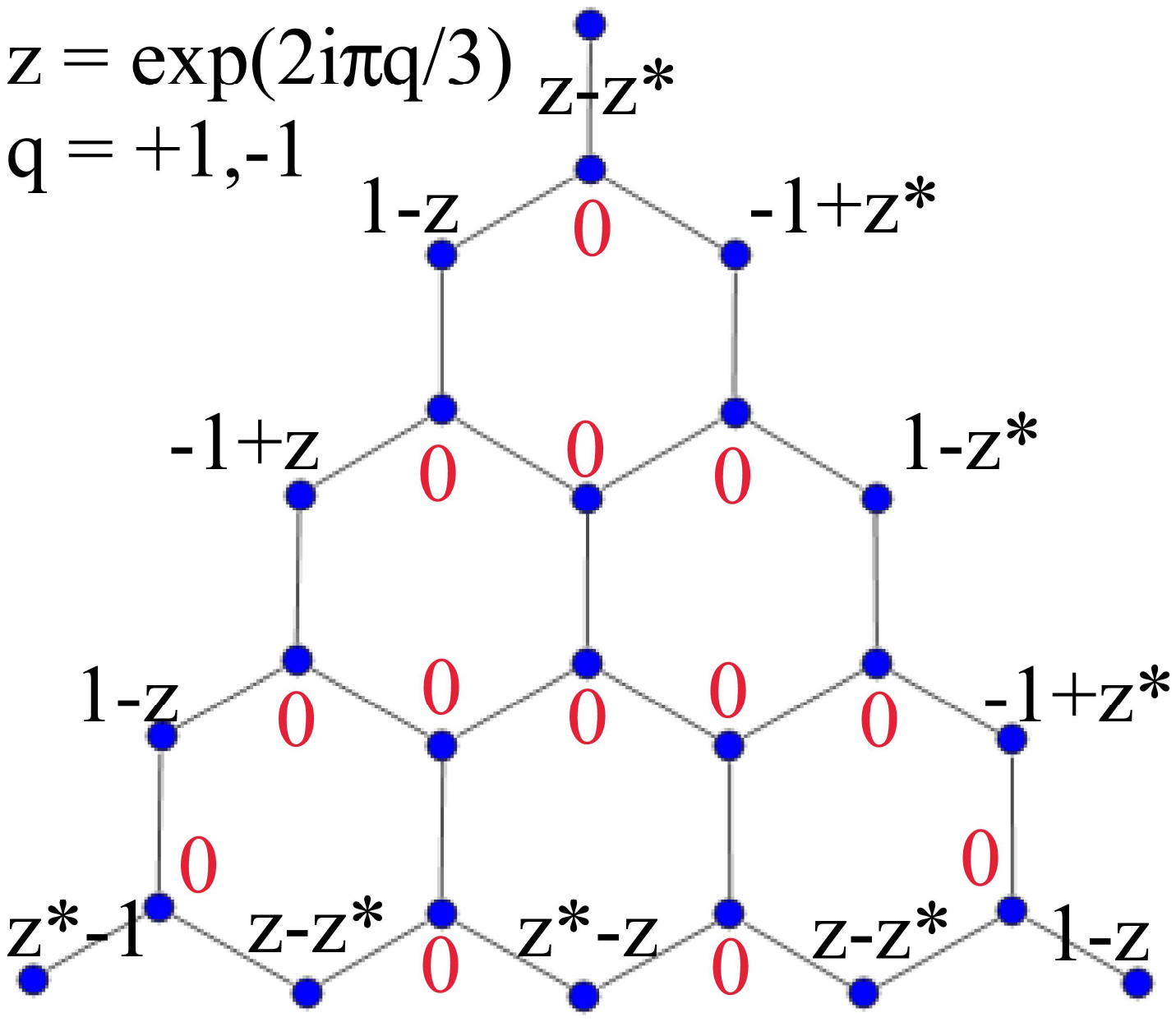}
     \caption{Edge states $\cet{1,1}$ and $\cet{1,-1}$ contain complex amplitudes in their sites. This is the so-called chiral representation.}
     \label{chiral}
\end{figure}

\begin{figure}[ht]
  \centering       \includegraphics[width=\linewidth]{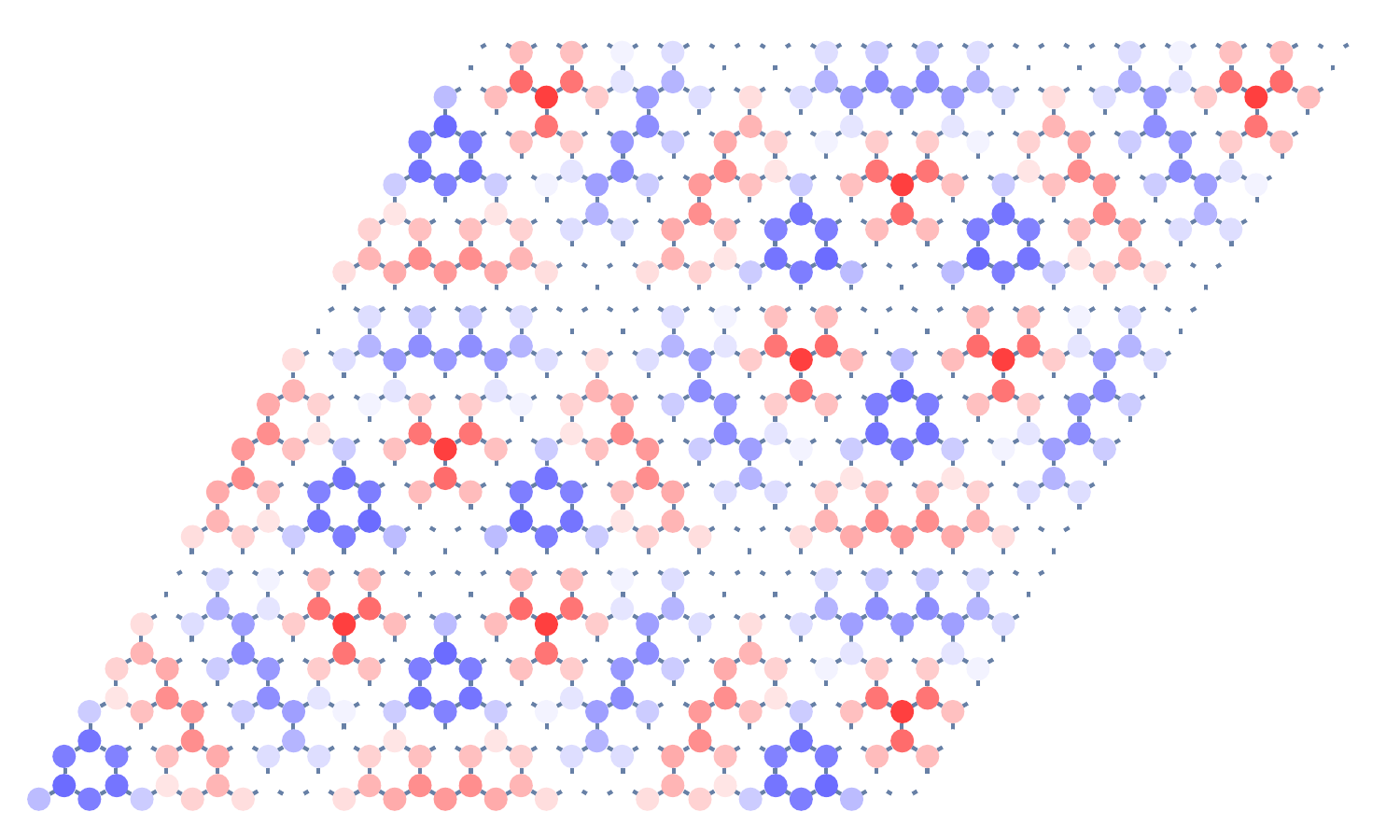}
     \caption{Triangular lattice built with barbed and zig-zag billiards. The colors represent a real eigenstate for $N=6$ as in fig. \ref{repre}.}
     \label{triangandbarbed}
\end{figure}

\begin{itemize}
	\item[1] If $3N_L=N-1$, we can generate the last 3 edge states using (\ref{non ort edge states}).
	\item[2] If $3N_L=N+1$, the innermost layer of the triangulene molecule consists only of a fixed point, i.e., a site not affected by rotations (see fig. \ref{central}) and from (\ref{Urotation}) we can infer that $\ket{3N_L,-1}$ and $\ket{3N_L,1}$ have a vanishing amplitude at the fixed point, therefore, both of these states have no amplitude in the innermost layer and as shown in the previous paragraph, they can be written in terms of the previous $\ket{l,q}$. After eliminating them from the generative process, we obtain $N-1$ as the number of edge states in our basis.
	\item[3] If $3N_L=N$, the innermost layer of state $\ket{3N_L,0}$ is the sum of the antisymmetric solution of the \linebreak $N=3$ triangulene plus its corresponding rotations (see Figs. \ref{N2layer}a-\ref{N2layer}c), therefore, their amplitudes at this layer vanish and must be eliminated from the generative process, which gives us once more $N-1$ as the number of edge states in our basis.
\end{itemize}

Lastly, we can implement a simple Gram–Schmidt process to achieve full orthogonality (note the round bracket in our ket)
\begin{align}
&\cet{1,q}=\ket{1,q}/\sqrt{\braket{l,q|l,q}},\quad q=-1,0,1\\
\label{GS}
&\cet{l,q}=\frac{\ket{l,q}-\sum_{k=1}^{l-1}\langle k,q|l,q)\ket{k,q}}{\sqrt{\braket{l,q|l,q}-\sum_{k=1}^{l-1}|\langle k,q|l,q)|^2}},\quad\begin{split}l=2,\dots,N_L,\\q=-1,0,1.\end{split}
\end{align}
For ease in the visualization, we display the real and imaginary parts, which are degenerate by time-reversal symmetry. See fig. \ref{ort ed state} for the $N=6$ billiard resembling triangulene. The state $\cet{1,-1}$ is already real. The plots correspond to our seed states in figs. \ref{bothlayers}a-\ref{bothlayers}b. Additionally, in fig. \ref{chiral} we show the full edge states $\cet{1,-1}$ and $\cet{1,1}$, indicating their corresponding complex amplitudes. These states carry a current, understood as a variation in the phase; the chirality of these states is parameterized by $q=\pm1$.\\



\subsection{
Completeness of solutions \label{completeness}}

As discussed previously, we must have a total of \linebreak $(N-1)^2$ l.i. solutions. Now that we have an account of all the possible eigenstates, we must verify that this number is indeed correct by incorporating the number of periodic solutions previously obtained. From (\ref{k quantx}) and (\ref{k quanty}) we obtain an infinite lattice of points that relate to valid solutions. In order to obtain only the l.i. solutions, we restrict to the first triangular sector in the Brillouin zone of fig. \ref{smallt}, which translates into the limits (\ref{sector qp}). In terms of integers $q$ and $p$, these limits are written in terms of floor functions as
\be
\label{dis limits}
1\leq p\leq\left\lfloor\frac{N-1}2\right\rfloor,\quad p\leq q\leq N-1-p.
\ee
Now we note that according to (\ref{states c}), for the $N_I$ internal points that satisfy $q\neq p$, each one leads to 4 l.i. solutions ($\psi_1$, $\psi_2$ are two-fold degenerate, plus the $\pm$ energy selection), while the number $N_A$ of points along a symmetry axis that satisfy $q=p$ (green line in fig. \ref{smallt}) only yield 2 l.i. solutions ($\psi_1=0$). These observations help us to generate the spectrum of the billiard with the correct considerations for degeneracies. We use (\ref{dis limits}) to determine $N_I$ (points inside) and $N_A$ (axial points):
\be
\begin{gathered}
N_I=\sum_{p=1}^{\left\lfloor\frac{N-1}2\right\rfloor}\sum_{q=p+1}^{N-1-p}1=\frac14(N-1)(N-2)-\frac12\left\lfloor\frac{N-1}2\right\rfloor,\\
N_A=\sum_{q=1}^{\left\lfloor\frac{N-1}2\right\rfloor}1=\left\lfloor\frac{N-1}2\right\rfloor.
\end{gathered}
\ee
As a consequence, the number $N_1$ of solutions with $E\neq0$ is
\be
N_1=4N_I+2N_A=(N-1)(N-2).
\ee
Previously, we established that the number $N_0$ of l.i. edge states is $N-1$, thus the total $N_T$ of l.i. solutions is, as expected:
\be
N_T=N_0+N_1=(N-1)^2.
\ee
As a closing remark, thanks to the Gram–Schmidt process employed in (\ref{GS}), we are able to write a full completeness relation for the triangulene molecule in the basis of eigenstates
\be
\begin{aligned}
\unit &=\sum_{E_n\neq0}\cet{n}\cra{n}+\sum_{E_n=0}\cet{n}\cra{n}\\
&=\sum_{n=1}^{(N-1)(N-2)}\cet{n,E_n}\cra{n,E_n}+\sum_{n=1}^{N-1}\cet{n,0}\cra{n,0}.
\end{aligned}
\ee

\section{Conclusions}

In this paper, we report analytical solutions for a triangular billiard made of graphene with zig-zag edges, for the first time. Our construction was divided into periodic solutions and edge states. Moreover, we show that the periodic solutions also constitute stationary waves for a complementary graphene flake made of barbed boundaries, as their images are comprised in an infinite triangular lattice. See fig. \ref{triangandbarbed} obtained from our solutions. Although the barbed edge states were not presented, it should be easy to extend our discussions to such cases. In the tradition of mathematical physics, reporting new exact solutions of the Schr\"odinger equation requires verification of completeness: This was done by directly counting the number of eigenstates in comparison with the molecular size. It is no surprise that the full set of solutions could be obtained for such a simplistic triangular figure, given the well-known integrability of its Helmholtz counterpart \cite{Pinsky,japanese,casati}. Other shapes are likely susceptible to similar treatments, such as hexagons or parallelograms, but surprisingly not rectangles, as they combine different types of boundaries. Recently, Heller et al. \cite{Heller2} were able to observe signatures of quantum scars \cite{Heller1} in irregular shapes made of graphene, but the corresponding mathematical description cannot coincide with predictions for elliptic operators. The use of Dirac operators, instead, should help to settle the issue, with the exception of \textit{finite} tight-binding models, for which edge states make their surprising appearance.

Regarding such degenerate solutions at $E=0$, we have been successful in classifying all edge states employing the $C_{3v}$ symmetry of triangulene in addition to an extra orthogonalization step. It is very important to recognize that these solutions differ from the conventional Dirac point solution of infinite graphene. Edge states have been important in recent work due to their applications in topological insulators as well as Majorana modes \cite{majorana}. The possibility of producing rotated layers from this kind of sheets is still open; see e.g. \textit{moir\'e} materials \cite{moire} and, possibly, their molecular counterparts. Further explorations will be left for the future, as they require a careful determination of the inter-layer electronic coupling. Finally, we note that these \textit{moir\'e} patterns are useful in the variation of the effective wavelength for propagating electrons in infinite media, but in finite media, such effects are yet to be analyzed.\\

\section*{Acknowledgments}

Financial support from CONAHCYT under Grant No. CF-2023-G-763 is acknowledged. D.C. is grateful to VIEP-BUAP for support under Project No. 100518931-BUAP-CA-289.

\appendix

\section{Trigonometric Diophantine equation}
\label{diof}
Firstly, we define the following abbreviations
\be
\label{arg def}
\begin{gathered}
k_1=\mb k\cdot\mb a_1,\quad k_2=\mb k\cdot\mb a_2,\\
\chi_1=-nk_1+Nk_1/2+\alpha,\quad\theta_1=(N/2)(k_1-2k_2)\\
\chi_2=nk_2-Nk_2/2+\alpha,\quad\theta_2=(N/2)(k_2-2k_1)\\
\chi_3=-\chi_1-\chi_2+3\alpha,\quad\theta_3=-\theta_1-\theta_2,    
\end{gathered}
\ee
now we write (\ref{condition for 12}) explicitly
\be
\begin{split}
&0=\sin(-nk_1+N(k_1-k_2)+\alpha)+\sin(nk_1-Nk_2-\alpha)\\
&+\sin(nk_2-Nk_1+\alpha)+\sin(-nk_2-N(k_1-k_2)-\alpha)\\
&+\sin(n(k_1-k_2)+Nk_2+\alpha)+\sin(-n(k_1-k_2)+Nk_1-\alpha)\\
&=\sum_{i=1}^3\left[\sin(\theta_i+\chi_i)+\sin(\theta_i-\chi_i)\right]=2\sum_{i=1}^3\sin\theta_i\cos\chi_i,
\end{split}
\ee
where in the last equality a trigonometric identity was used. Since this must be true for all values of $n$, we have the following restriction $\sin\theta_i=0$, $i=1,2,3$. This is achieved by the requirement
\be
\label{thetas}
\theta_1=-\pi p,\quad\theta_2=-\pi q,
\ee
where $q$ and $p$ are integers and for convenience, are written with a minus sign. Now we substitute (\ref{thetas}) into (\ref{arg def})
\be
(N/2)(k_1-2k_2)=-\pi p,\quad (N/2)(k_2-2k_1)=-\pi q.
\ee
This is a system of equations for $k_1$ and $k_2$, once we solve it, we obtain (\ref{k quantx}) and (\ref{k quanty}):
\begin{gather}
\label{k quantx2}
k_1=\mb k \cdot\mb a_1=2 \pi(2q+p)/ 3 N,\\
\label{k quanty2}
k_2=\mb k \cdot\mb a_2=2 \pi(2p+q)/ 3 N.
\end{gather}

\bibliography{mybib}

\end{document}